\documentclass[fleqn,usenatbib]{mnras}
\pdfoutput=1
\usepackage{newtxtext}
\usepackage{mathptmx}
\usepackage[T1]{fontenc}
\DeclareRobustCommand{\VAN}[3]{#2}
\let\VANthebibliography\thebibliography
\def\thebibliography{\DeclareRobustCommand{\VAN}[3]{##3}\VANthebibliography}
\defcitealias{Dhar:surfden}{DW10}
\defcitealias{Dhar:ellipticals}{DW12}
\defcitealias{Retana:analprops}{RvHGBF12}
\defcitealias{Nav04:einasto}{N04}
\usepackage{graphicx}	
\usepackage{amsmath}	
\usepackage{amssymb}	
\usepackage{booktabs,caption}
\usepackage[flushleft]{threeparttable}
\title[Projected Mass \& Surface Density of Einasto profiles]{High-accuracy analytical solutions for the Projected Mass (Counts) \& Surface Density (Brightness) of Einasto profiles}
\author[B. K. Dhar]{
Barun K. Dhar,$^{1}$\thanks{E-mail: bdhar@ucsc.edu}\\
$^{1}$ Physics Department, University of California Santa Cruz, Santa Cruz, CA 95064\\
}
\date{Accepted 2021 March 31. Received 2021 March 31; in original form 2021 February 16.}
\pubyear{\href{https://doi.org/10.1093/mnras/stab1029}{2021}}
\begin{document}
\label{firstpage}
\pagerange{4583--4588}
\maketitle
\begin{abstract}
The Einasto profile has been successful in describing the density profiles of dark matter haloes in $\Lambda$CDM {\it N\it}-body simulations. It has also been able to describe multiple components in the surface brightness profiles of galaxies. However, analytically projecting it to calculate  quantities under projection is challenging. In this paper, we will see the development of a highly accurate analytical approximation for the mass (or counts) enclosed in an infinitely long cylindrical column for Einasto profiles--also known as the projected mass (or counts)--using a novel methodology. We will then develop a self-consistent high-accuracy model for the surface density from the expression for the projected mass. Both models are quite accurate for a broad family of functions, with a shape parameter $\alpha$ varying by a factor of 100 in the range $0.05 \lesssim \alpha \lesssim 5.0$, with fractional errors $\sim 10^{-6}$ for $\alpha \lesssim 0.4$. Profiles with $\alpha \lesssim 0.4$ have been shown to fit the density profiles of dark matter haloes in {\it N\it}-body simulations as well as the luminosity profiles of the outer components of massive galaxies. Since the projected mass and the surface density are used in gravitational lensing, I will illustrate how these models facilitate (for the first time) analytical computation of several quantities of interest in lensing due to Einasto profiles. The models, however, are not limited to lensing and apply to similar quantities under projection, such as the projected luminosity, the projected (columnar) number counts and the projected density or the surface brightness.\\
\textcolor{magenta}{{\bf Note\bf}: This is a pre-copyedited, author-produced PDF of an article accepted for publication in MNRAS following peer review. In the published version available \href{https://doi.org/10.1093/mnras/stab1029}{here}, the value of the constant $c_0$ is listed incorrectly in Table 1 (corrected in an \href{http://dx.doi.org/10.1093/mnras/stab1760}{Erratum} and in this Arxiv version). This mistake does not affect the results in the paper for which the correct value have been used. }.
\end{abstract}
\begin{keywords}
dark matter -- galaxies: clusters -- galaxies: structure -- gravitational lensing: strong --  methods: analytical -- methods: numerical. 
\end{keywords}
\section{Introduction}\label{intro}
The Einasto profile is a 3-parameter function that is used to describe the spherically averaged 3-dimensional spatial density $\rho(r)$ of galaxies and dark matter haloes. The function is of the form:
\begin{align}\label{modexp1}
\rho(r)=\rho_s \ e^{- b [(r/r_{s})^{\alpha}-1 ]} = \rho_0 \ e^{- b (r/r_s)^{\alpha}}
\end{align}
where, $\rho_s$ is the 3D volume density at a scale-length $r_s$, $\rho_0=\rho_s e^b$ is the central density with $b=2/\alpha$ and $\alpha$ is the parameter defining the shape of the profile. The exponential ($\alpha=1$) and the gaussian ($\alpha=2$) functions are members of this family of functions. The function is also described using a shape-parameter $n$ where $n=1/\alpha$. Throughout this paper, we will use the form in equation \eqref{modexp1}.

This functional form, first proposed by~\citet{Einasto:1965}, has been used to model multiple components in the structure of spiral and elliptical galaxies~\citet*{Rummel:onm31, Tenjes:onm87, Tempel:4594} and \citealt{Dhar:ellipticals} (hereafter \citetalias{Dhar:ellipticals}). It has also been successful in describing the spherically averaged 3D density profiles of dark matter haloes in N-body simulations~\citep{Nav04:einasto, Merritt:05, Merritt:06}. This is probably because the Einasto profile is a very good approximation to the density profile $\rho(r)$ of DARKexp---an equilibrium statistical mechanical theory of collisionless self-gravitational systems where $\rho(r)$ does not have an analytical form ~\citep{HjorthW:10, HjorthW:15}. This similarity with DARKexp has been observed in N-body simulations \citep*{WilliamsH:10} as well as in real clusters of galaxies \citep*{Beraldo:13}.

In imaging studies, observations are of quantities projected in the 2-dimensional (2D) plane of the sky at a projected radial distance $R$, such as the surface brightness $\Sigma(R)$. Likewise, projected quantities such as the convergence $\kappa (R)$ and the mass enclosed $M_{2D}(R)$ in an infinitely long cylindrical column of radius $R$ are important quantities of interest in studying the effects of gravitational lensing. All of these quantities are related to a projection of the volume density $\rho(r)$.

Nevertheless, except for the case of $\alpha=2$, analytically projecting \autoref{modexp1} {\it and calculating} $\Sigma(R)$, $\kappa(R)$ and $M_{2D}(R)$ exactly is impossible (see \citealt{Dhar:surfden} (hereafter, \citetalias{Dhar:surfden}) and \citealt*{Cardone:2005}). \citetalias{Dhar:surfden} developed an analytical approximation for $\Sigma(R)$ and \citealt{Retana:analprops} (hereafter, \citetalias{Retana:analprops}) used the {\it Fox-H} function to {\it describe} exact solutions for several quantities. However, save for the case of the exponential ($\alpha=1$) and the gaussian ($\alpha=2$), the solutions in \citetalias{Retana:analprops} cannot be {\it computed} using standard special functions. Moreover, these two profiles do not cover several systems of astrophysical interest. 

\begin{figure}
\includegraphics[width=\columnwidth]{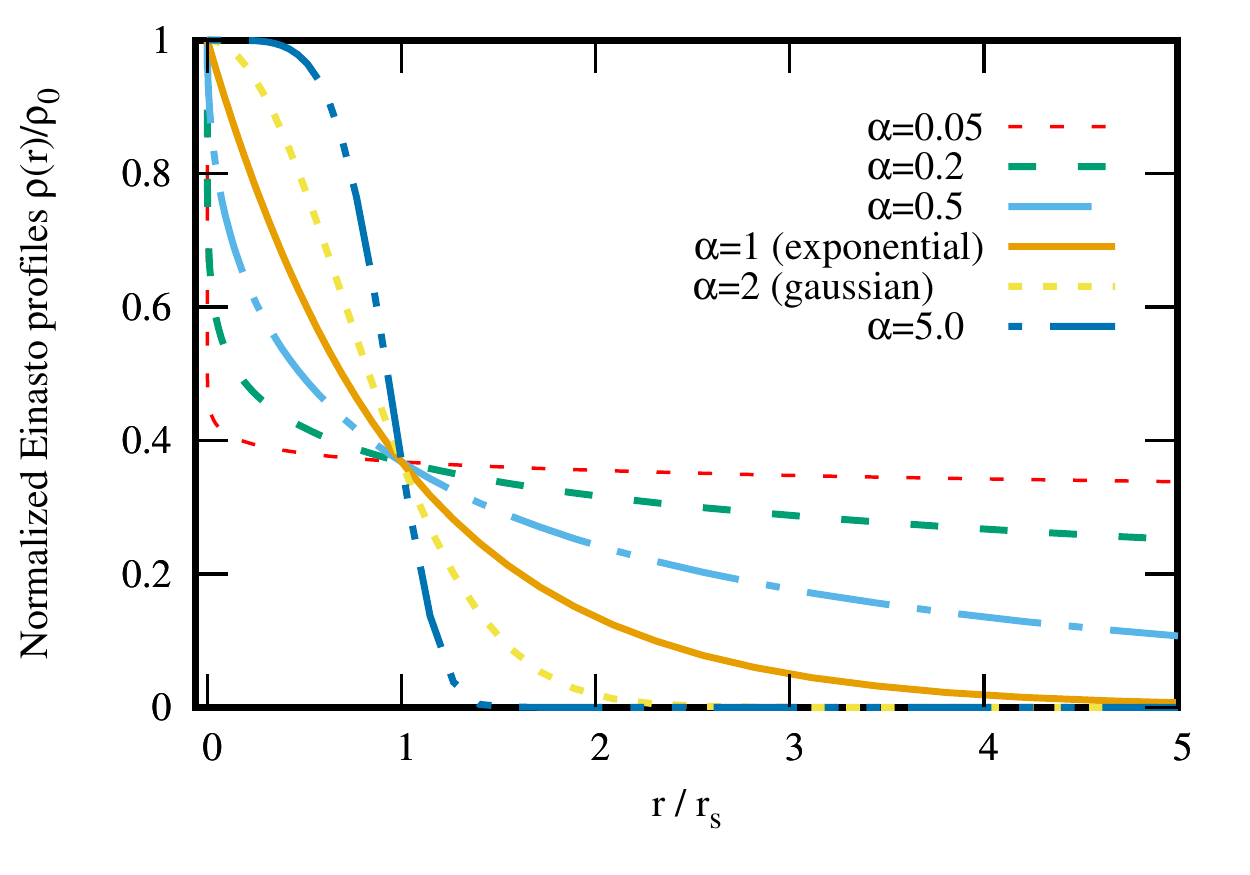}
\vspace{-5mm}
\caption{Normalized Einasto profiles of the form $\rho(r)/\rho_0 = e^{-(r/r_s)^\alpha}$ where $\alpha$ controls the shape of the function. Profiles with $\alpha > 1$ can be used to model systems where the density rises gradually as $r \to 0$ but falls off sharply at larger $r$, while profiles with $\alpha < 1$ apply to systems where the density rises sharply as $r \to 0$ but falls off gradually at larger $r$.}
\label{Einastoprofiles}
\end{figure}

For instance, \citet{Dhar:ellipticals} have shown that the outer (dominant) component of massive ellipticals in the Virgo Cluster are described by Einasto profiles with $0.1 \lesssim \alpha \lesssim 0.2$ and other components in those galaxies as well as all components in smaller cuspy ellipticals have profiles with $0.2 \lesssim \alpha \lesssim 1.5$. The authors note the striking similarity in the range of $\alpha$ seen in the outer components of massive ellipticals and the range seen in the N-body simulations and conjecture that the surface brightness profiles of the outer components may be reflecting the shape of the underlying dark matter halo. \citet{GaoL:08} and \citet{Klypin:multidark} explored the evolution of dark matter haloes (in N-body simulations) as a function of redshift and found similar trends---$0.12 \lesssim \alpha \lesssim 0.3$ at low ($z<1$) redshift, increasing to $0.25 \lesssim \alpha \lesssim 0.44$ at high ($z>3$) redshift. We will therefore, hereafter, refer to $0.1 \lesssim \alpha \lesssim 2.0$ as the {\it range of current practical interest}. The results of this paper have nevertheless been tested for $0.05 \leq \alpha \leq 5.0$ to ensure stability of the solutions outside the range of current interest. As shown in~\autoref{Einastoprofiles},  this range covers a broad family of profile shapes as elaborated in the caption of ~\autoref{Einastoprofiles}.

While an analytical approximation to calculate the surface density $\Sigma(R)$ for Einasto profiles has been developed in \citetalias{Dhar:surfden}, a solution for calculating the projected mass (or counts), using existing practice (discussed in \autoref{newmethod}) has remained elusive. The lack of the latter limits our ability to use projected mass (or counts) to study Einasto profiles using observed data. For example, while studying flux ratios in the gravitationally lensed system HS 0810+2554 ~\citep{Jones:gravlens}, the authors could not calculate the projected mass and the Einstein radius (due to an Einasto profile for the lens) by either using the results from \citetalias{Retana:analprops}  or from \citetalias{Dhar:surfden} and had to use a S{\'e}rsic equivalent profile from \citetalias{Dhar:surfden} to calculate the Einstein radius. However, such equivalences are not accurate as ~\citetalias{Dhar:surfden} notes. The goal of this paper is therefore twofold:
\begin{enumerate}
\item  First is to present a new method (in ~\autoref{newmethod}) for calculating the projected mass (or counts) in a cylindrical column of infinite length. We will refer to this quantity as $M_{2D}(R)$; 
\item  Second is to show how this new method can be applied to develop an analytical solution for the projected mass (or counts) of Einasto profiles, and then use that expression to develop a solution for the surface density $\Sigma(R)$, as we will see in ~\autoref{approxderiv}.
\end{enumerate}

In that light, for the rest of this paper we will use the phrase {\it analytically integrable} to imply integrable in closed form using functions in standard \texttt{C++} math libraries such as the {\it GNU Scientific Library (GSL)} \citep{GSL:lib} so that the model can be used in a broad range of computing environments. We will also use the notation of $M_{2D}(R)$ to represent any integrated volume density under projection. We will therefore refer to this quantity either as the {\it columnar mass} or the {\it projected mass} or the {\it projected luminosity} or as the {\it columnar number counts}. Likewise, $\Sigma(R)$ will refer to either a surface (or projected) mass density or the projected number density or a projected luminosity density, also known as the surface brightness.
\vspace{-5mm}
\section{A New Method for calculating the Mass (or Counts) in a cylindrical column $M_{2D}(R)$:}\label{newmethod}
\begin{figure}
\includegraphics[width=\columnwidth]{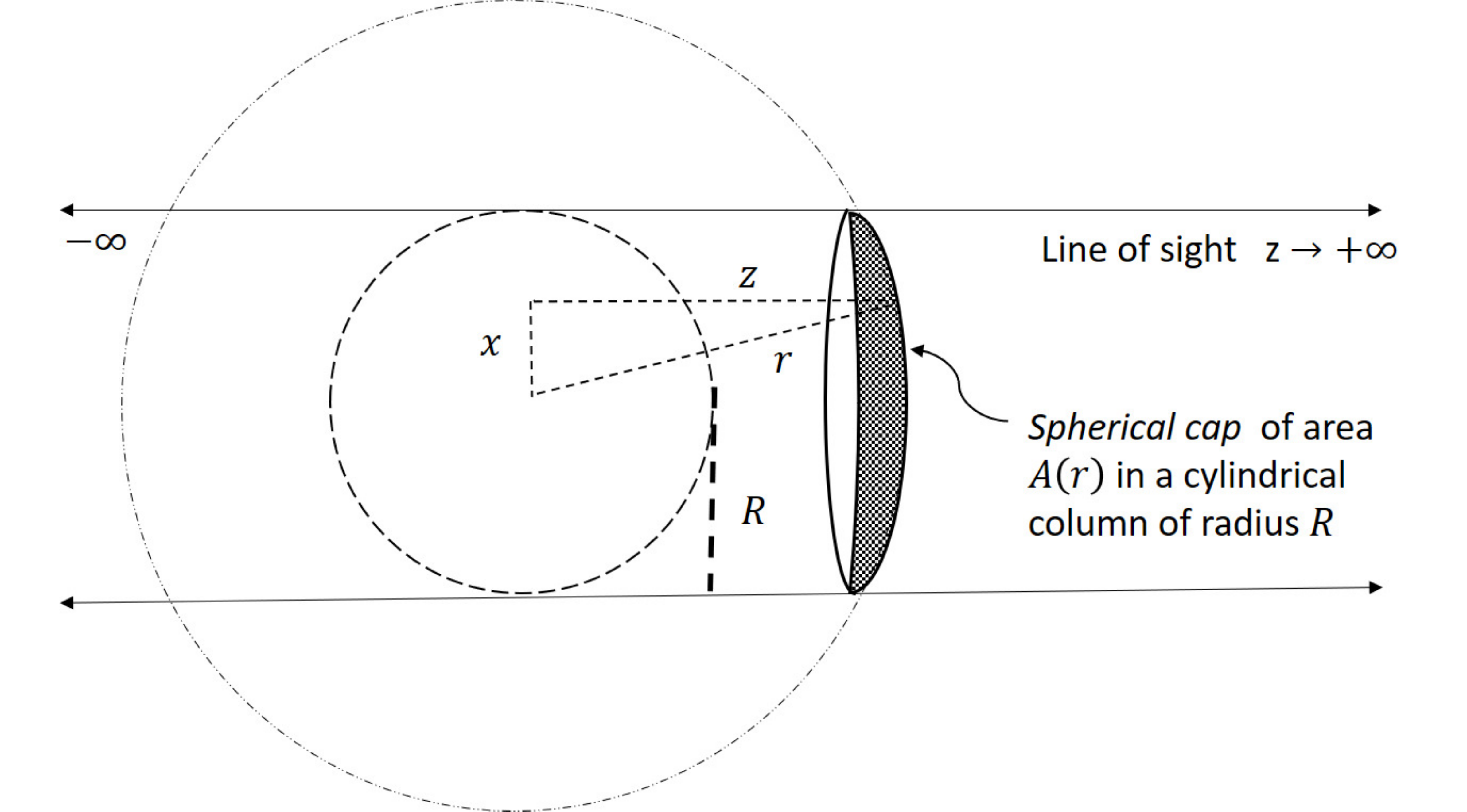}
\caption{Schematic showing $x, r, z$ and $R$ in a cylindrical column of infinite length and radius $R$. The shaded region is the spherical cap of surface area $A(r)$ at $r$ with a base radius $R$.}
\label{SphCapfig}
\end{figure}

It is a common practice to calculate the mass (or counts) within an infinitely long cylindrical column of radius $R$,  $M_{2D}(R)$, by integrating the projected surface density $\Sigma(R)$ over the projected radius $R$:
\begin{flalign}\label{TradCp}
M_{2D}(R) &= 2\pi \int^{R}_{0} x \ \Sigma(x) \ dx
\end{flalign}
We can then relate $M_{2D}(R)$ to the intrinsic density $\rho(r)$ since $\Sigma(R)$ is related to $\rho(r)$ through the forward Abel transform of $\rho(r)$:
\begin{flalign}\label{AbelTransform}
\Sigma(R)=2 \int_{0}^{\infty} \rho(r) dz  \  \  OR  \ \ 2 \int_R^\infty \frac{\rho(r) \ r}{\sqrt{r^2-R^2}} \ dr 
\end{flalign}
where, $z=\sqrt{r^2-R^2}$ is a coordinate along the line of sight to the observer.  Refer to \autoref{SphCapfig} for a schematic showing $z$, $r$ and $R$.
Hence, analytically computing $M_{2D}(R)$ requires computing a double integral---one over intrinsic space and one over projected space:
\begin{flalign}\label{C2Ddouble}
M_{2D} (R) &=4\pi \int^{R}_{0} \ x  \ \left( \ \int^{\infty}_{x}  \frac{\rho(r)}{\sqrt{1-(\frac{x}{r})^{2}}} \ dr \right) \ \ dx 
\end{flalign}
Analytical computation of $M_{2D}(R)$ becomes impossible if either one of the integrals is not analytically computable. For the Einasto profile, \citetalias{Retana:analprops} presents an analytical {\it description} using the {\it Fox-H} function, but as of now there is no way to {\it compute} {\it Fox-H} functions.

We will therefore develop an analytical approximation for $M_{2D}(R)$ inspired by a novel approach first proposed by \citet{Dhar:thesis} but build the solution differently. We recall the proposal in \citet{Dhar:thesis} that since the integrals of $\rho(r)$ and $\rho(r) \ r^2$ exist in the domain $0 \leq r \leq \infty$ for the Einasto profile, we can calculate $M_{2D}(R)$ as a sum of the intrinsic 3D mass $M_{3D}(r)$ within a sphere of intrinsic radius $r=R$ and the mass over spherical caps of base radius $R$ and surface area $A(r)$ (see schematic in \autoref{SphCapfig}), instead of the conventional approach of equations \eqref{TradCp} and \eqref{C2Ddouble}. This significantly simplifies the calculation of $M_{2D}(R)$ from a double integral over two domains ($r$ and $R$) as in equation \eqref{C2Ddouble} to a single integral over just the intrinsic space $r$ as in:
\begin{align}\label{newCp}
M_{2D} (R)= M_{3D}(R)+ 2 \int^{\infty}_{R} \rho(r) A(r) dr
\end{align}
\begin{align}\label{Cr3D}
where, ~~~ M_{3D}(R) &=4 \pi \int^{R}_{0} r^{2} \ \rho(r) \ dr
\end{align}
is the mass enclosed in a sphere of radius $r=R$, \ and
\begin{align}\label{acap}
  A(r)=2 \pi r^{2} \left(1-\sqrt{1-(R/r)^{2}}\right) 
\end{align}
is the area of a spherical cap at an intrinsic radial distance $r$ with $R$ as the radius of the base of the spherical cap (see \autoref{SphCapfig}).
Expanding the square root in \autoref{acap}, we get:
\begin{equation}\label{CpPowerorig}
\begin{split}
& M_{2D}(R)=M_{3D}(R) \ +  \\
& 4 \pi \int^{\infty}_{R} \rho(r) \ r^{2} \left(\frac{1}{2} \frac{R^2}{r^2} + \frac{1}{8} \frac{R^4}{r^4} + \frac{1}{16} \frac{R^6}{r^6} + \frac{5}{128} \frac{R^8}{r^8}+.... \right) \ dr
\end{split}
\end{equation}
which can be rewritten as 
\begin{equation}\label{CpPower}
\begin{split}
& M_{2D}(R)=M_{3D}(R) + 2 \pi R^{2} \int^{\infty}_{R} \rho(r) \ dr \ + \\
& 4 \pi R^{2} \int^{\infty}_{R} \rho(r)  \left(\frac{1}{8} \frac{R^2}{r^2} + \frac{1}{16} \frac{R^4}{r^4} + \frac{5}{128} \frac{R^6}{r^6} +.... \right) \ dr
\end{split}
\end{equation}
The power series in~\autoref{CpPower} is convergent for all $r \geq R$: as $r \to R$ the expansion tends to 1 and as $r \to \infty$ the expansion approaches $\frac{1}{2}$. This opens up the possibility of developing closed form analytical approximations for $M_{2D}(R)$ by limiting the series to a few terms.

We proceed to do so for the Einasto profile (\autoref{modexp1}) by observing that $M_{3D}(r)$ and the first integral in \autoref{CpPower} are analytically integrable, while the remaining terms are not integrable in terms of functions in standard math libraries. For those terms, we exploit a simplification due to Bonnet's $2^{nd}$ mean value theorem (MVT) which states that (see \citet{Bonnet:2ndmvt} and \citet{Hobson:2ndmvt})---%

{\it if f(x) is a finite and monotonic function in the interval (a,b) and g(x) possesses a Lebesgue integral---i.e. is integrable in (a, b) or has at most one non-absolutely convergent improper integral in (a, b)---then there exists a point $\zeta$, in $a \leq \zeta \leq b$,
such that, 
\begin{flalign}
\int_a^b f(x) g(x) dx = f(a) \int_a^\zeta g(x) dx + f(b) \int_\zeta^b g(x) dx
\end{flalign}
\it}
We observe that for each term in the power series expansion (\autoref{CpPower}) we can identify terms like $(R/r)^m$ with $f(x)$ and $\rho(r)$ with $g(x)$. Then, by Bonnet's $2^{nd} MVT$ , there exists a $\zeta_i >1$ per term 
\begin{align}\label{betai}
such~~that~,~~~\int_{R}^{\infty} \beta_i \ \Big(\frac{R}{r}\Big)^{m} \ \rho(r) dr = \ \beta_i \ \int_{R}^{\zeta_i R} \rho(r) dr
\end{align}
where, $\beta_i$ are the coefficients of the power series expansion.

By doing so for each of the remaining terms of the expansion we can make those terms analytically integrable since $\rho(r)$ as equation~\eqref{modexp1} is analytically integrable. The values for $\zeta_i$, however, can be a function of both $R$ and $\alpha$. Additionally, the number of terms of the expansion that would be needed to develop a good approximation may also vary with $\alpha$.  This leads to a complicated model. 

In \autoref{approxderiv}, we will discuss the conditions under which we can consider each $\zeta_i$ a constant (independent of $\alpha$) thereby allowing us to obtain a simplified (yet highly accurate) model. Before doing so, we will evaluate the accuracy of the numerical integrations (in \autoref{accuracygauss}).
%
\section{Accuracy of numerical integrations}\label{accuracygauss}
In order to estimate the $\zeta_i$, we first construct numerically integrated profiles for $M_{2D}(R)$ for several values of $\alpha$ in the range $0.05 \leq \alpha \leq 5.00$, corresponding to an Einasto-index $n$ in the range $0.2 \leq n \leq 20.0$---a factor of 100. Since closed form analytical expressions for $\Sigma(R)$ and $M_{2D}(R)$ (using equations \eqref{AbelTransform} and \eqref{TradCp}) exist only for the case for the gaussian profile $\alpha=2.0$ (and at $R=0$ for all $\alpha$), we test the accuracy of the numerical integration routines using this case.

We observe that for all $\alpha$, \autoref{AbelTransform} (with $R=0$) gives us: 
\begin{flalign}\label{sigma0}
\Sigma (0)=\frac{2r_s \rho_0}{\alpha \ b^{1/\alpha}} \ \Gamma \left(\frac{1}{\alpha}\right) \ \ and  \ \ M_{2D} (0)=0
\end{flalign}
For $\alpha=2$ in \autoref{AbelTransform} we get: 
\begin{flalign}\label{sig2exact}
\Sigma(R)=\rho_0\ r_s \ \sqrt{\pi}  \ e^{-(R/r_{s})^2} 
\end{flalign}
which when used in \autoref{TradCp} gives us
\begin{flalign}\label{C2exact}
M_{2D} (R)=\rho_0 \ r_s^3 \  \pi^{3/2} \Big(1-e^{-(R/r_{s})^2}\Big)
\end{flalign}
We therefore use equations \eqref{sig2exact} and \eqref{C2exact} to test the level of accuracy of the numerical integration routines in computing $\Sigma_N$ and $M_{2D \ N} (R)$ for the case of $\alpha=2.0$. The fractional (or relative) errors of the numerical integrations---compared to the exact analytical expressions in equations \eqref{sig2exact} and \eqref{C2exact}---turn out to be $\sim 10^{-15}$ for $\Sigma_N (R) $  and $\sim 10^{-13}$ for $M_{2D N} (R)$ over a large dynamic range of $R/r_s$ covering $10^{-23}$ \% to $99.99$ \% of the total mass (\autoref{cumgauss} below). 

The mass enclosed within a sphere of intrinsic (3D) radius $r$ can be analytically calculated exactly (for all $\alpha$) using 
\begin{flalign}
M_{3D}(r)=4\pi \int_{0}^{r} \rho(r') \ r'^2 \ dr' = 4\pi \int_{0}^{r} \rho_0 \ e^{-b (r'/r_{s})^{\alpha}} \ r'^2 \ dr'
\end{flalign}
which gives us (with $b=2/\alpha$),  
\begin{flalign}
M_{3D}(r)=\frac{4\pi \rho_0}{\alpha} \frac{r_{s}^3}{b^{3/\alpha}} \ \gamma \Big(\frac{3}{\alpha}, b\Big(\frac{r}{r_s}\Big)^{\alpha}\Big) 
\end{flalign}%
This gives us the total mass ($r\to \infty$) for all $\alpha$ as:
\begin{flalign}\label{cum3Dtot}
M_{3D total}=\frac{4\pi \rho_0}{\alpha} \frac{r_{s}^3}{b^{3/\alpha}} \ \Gamma \Big(\frac{3}{\alpha}\Big) 
\end{flalign}
where, $\Gamma(a)$ is the gamma function and $\gamma(a,x)$ is the lower-incomplete gamma function.

For $\alpha=2.0$ (the gaussian), the above two equations become:
\begin{flalign}
M_{3D}(r)&= 2\pi \rho_0 \ r_{s}^3 \ \gamma \Big(\frac{3}{2}, \Big(\frac{r}{r_s}\Big)^{2}\Big) \ , \ and
\end{flalign}
\vspace{-3mm}
\begin{flalign}\label{cumgauss}
M_{3D total}= 2\pi \rho_0 \ r_{s}^3 \ \Gamma (1.5)
\end{flalign}
\section{Analytical Models for $M_{2D}(R)$ \& $\Sigma(R)$}\label{approxderiv}
Having generated several numerically integrated profiles for $\Sigma_N (R)$ and $M_{2DN} (R)$ for each $\alpha$, we need to determine---what can constitute a good enough approximation for our models for $M_{2D}(R)$? Since most profile measurement errors are of the order of a few percent, we define a much smaller level of fractional error $\sim 10^{-5}$ (or 0.001 \%) to be satisfied by our models for $M_{2D}(R)$.

From~\autoref{newmethod} we recall that, while the $\zeta_i$ in \autoref{betai} may depend on $\alpha$, our model would be simpler if we can make them independent of $\alpha$. This is not unreasonable to require for the Einasto profile since most of the $\alpha$ dependency of $M_{2D}(R)$ is already contained in the first two terms of \autoref{CpPower} which---i) are both integrable in closed form; and ii) make the largest contribution to $M_{2D}(R)$ for all $\alpha$. The rest of the terms like equation~\eqref{betai} make a small contribution in equation~\eqref{CpPower} to produce a better approximation than using just the first two terms. However, how many terms will be required can also vary with $\alpha$. We therefore replace the $\beta_i$ with weights $c_i$ and check to see if we can find a fixed number of terms that will give us an accuracy of the order of $10^{-5}$ or better for all $\alpha$. It turns out that if we use six terms of the type of \autoref{betai} in \autoref{CpPower}, with the $\beta_i$ replaced by weights $c_i$ for those terms, then the values of $c_i$ and $\zeta_i$ given in \autoref{coeff} ensures that we get solutions for $M_{2D}(R)$ at our desired level of $\sim 10^{-5}$ (or lower) across the board for all $\alpha$. We thus proceed to develop $M_{2D}(R)$ by first writing \autoref{CpPower} as:
\begin{align}\label{initial}
M_{2D}(R)=M_{3D}(R)+2\pi R^2 \Big( \int_{R}^{\infty} \rho(r) dr + \sum_{k=1}^{6} c_k \int_{R}^{\zeta_k R} \rho(r) dr \Big)
\end{align}
Defining the coefficients $c_0 =1.0-\sum_{k=1}^6 c_k$ and $\zeta_0=1.0$, we can rewrite the above expression as\footnote{Note the summation is now from 0 to 6 which helps simplify \autoref{Ap2dcount}.} 
\begin{align}\label{initial2}
M_{2D}(R)=M_{3D}(R)+2\pi R^2 \Big( 2 \int_{R}^{\infty} \rho(r) dr + \sum_{i=0}^{6} c_i \int_{R}^{\zeta_i R} \rho(r) dr \Big)
\end{align}
Since each term is now analytically integrable for the Einasto profile with $\rho(r)$ as \autoref{modexp1}, we get (with $b=2/\alpha$):
\begin{equation}\label{Ap2dcount}
\begin{split}
& M_{2D}(R)=\frac{4\pi \rho_0}{\alpha} \frac{r_{s}^3}{b^{3/\alpha}}\Bigg [ \gamma \Big(\frac{3}{\alpha}, b\Big(\frac{R}{r_s}\Big)^{\alpha}\Big) + \\ 
& \frac{b^{2/\alpha}}{2} \Big(\frac{R}{r_s}\Big)^2 \Bigg \{ 2\ \Gamma \Big(\frac{1}{\alpha}, b\Big(\frac{R}{r_s}\Big)^{\alpha}\Big) - \sum_{i=0}^6 c_i \Gamma \Big(\frac{1}{\alpha}, b\Big(\frac{\zeta_i R}{r_s}\Big)^{\alpha}\Big) \Bigg \} \Bigg ]
\end{split}
\end{equation}
where, $\Gamma(a,x)$ and $\gamma(a,x)$ are the incomplete gamma functions. \\

\begin{figure}
\includegraphics[width=\columnwidth]{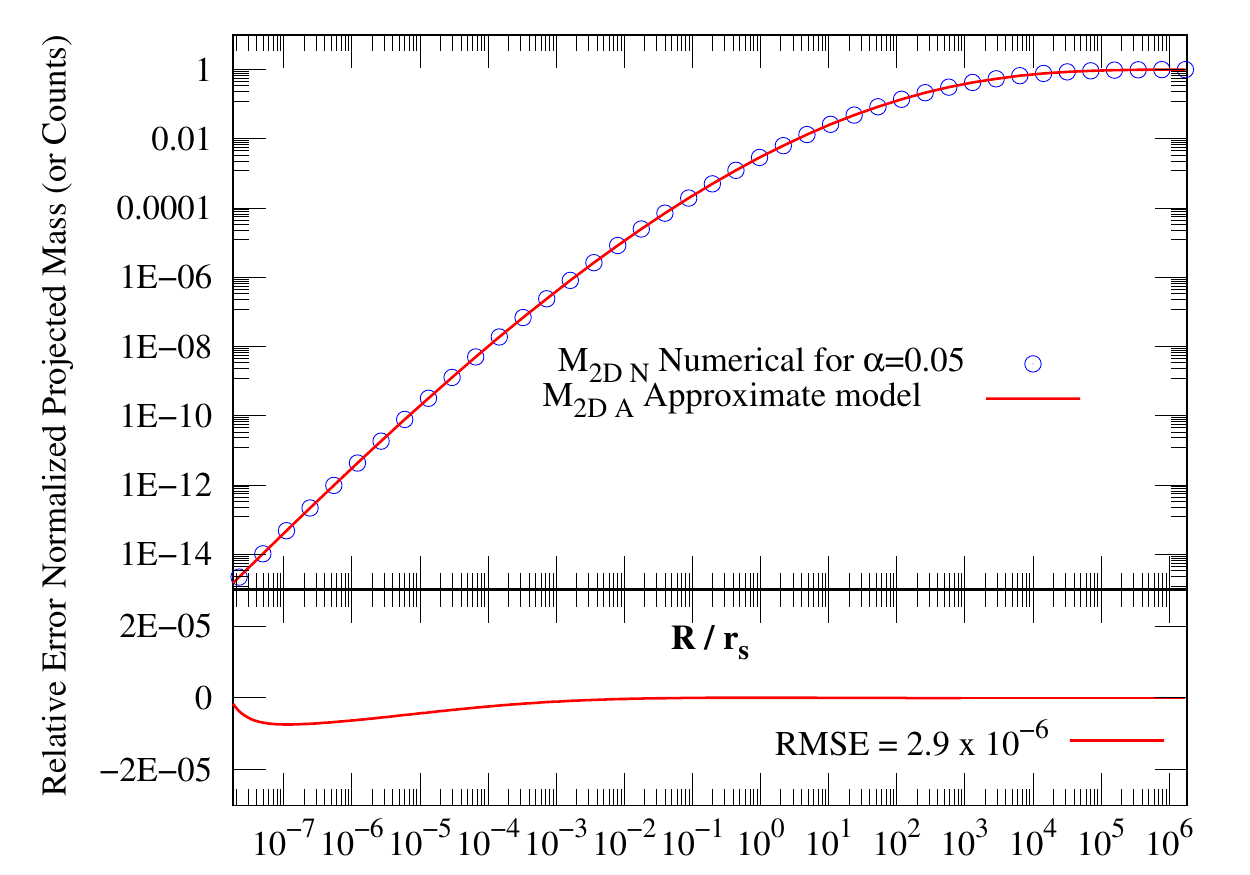}
\caption{Normalized Projected (Columnar) Mass (or counts) $M_{2D} (R)$ for an Einasto profile with $\alpha=0.05$ covering up to 99.5\% of the total mass (counts). The approximation retains excellent accuracy over 14 orders of magnitude in $R/r_s$ and over 15 orders of magnitude in $M_{2D}(R)$. Figures~\ref{MassAlpha005}--\ref{MassAlpha2} are normalized with respect to the total mass (or counts) $M_{3D total}$ (equation~\ref{cum3Dtot}) and Figures~\ref{SigAlpha05}--\ref{SigAlpha2} are normalized with respect to the central surface density $\Sigma_0$ (equation~\ref{sigma0}). The error panels show the relative error between the models derived in this paper and the numerically integrated values. The high accuracy of the approximation in these figures, over the entire radial extent, is more apparent when we compare them with profile measurement errors in observations and simulations which are typically $\sim 1-10\%$. }
\label{MassAlpha005}
\end{figure}

\begin{figure}
\includegraphics[width=\columnwidth]{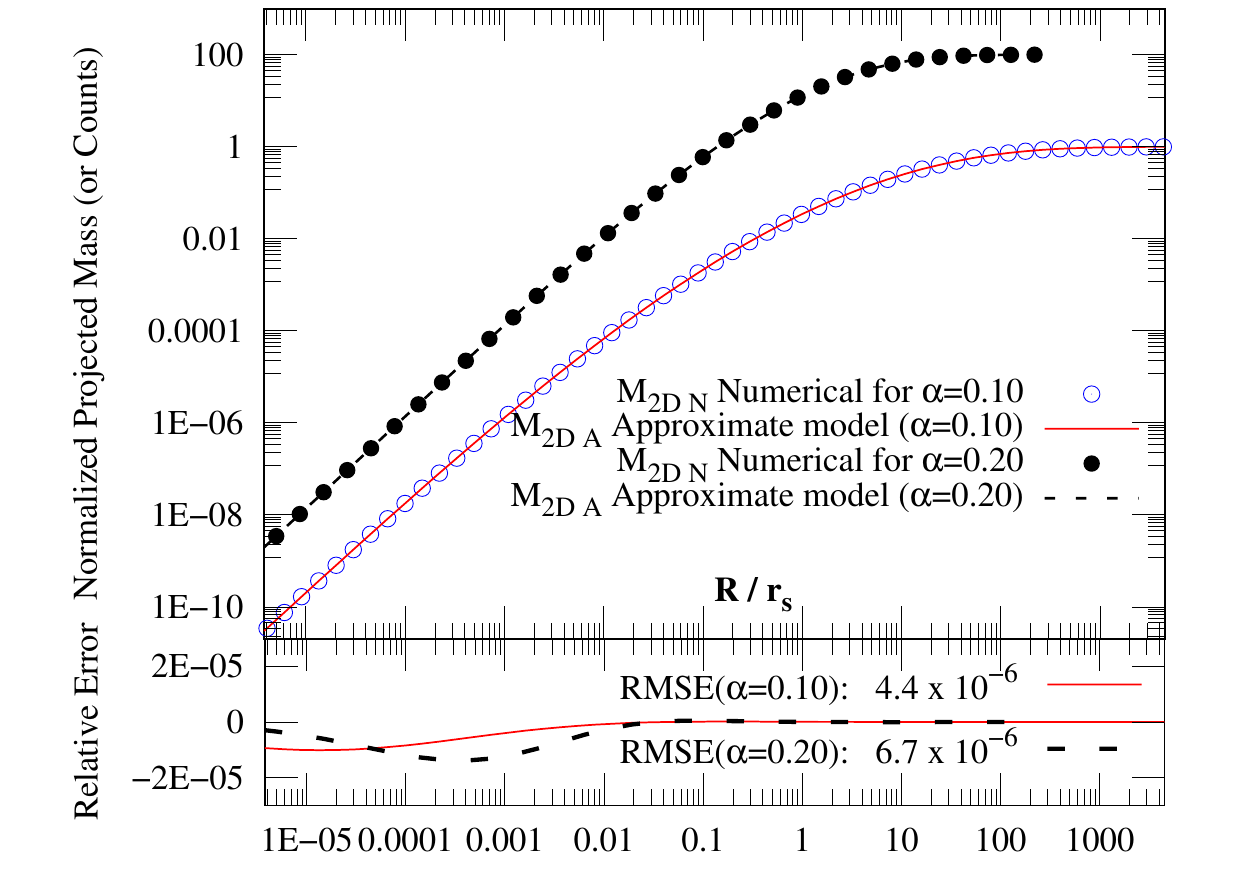}
\caption{Analytical models of normalized $M_{2D} (R)$ for Einasto profiles with $\alpha=0.10$ (solid) and with $\alpha=0.2$ (dashed), the latter scaled up by a factor of 100. Both profiles extend up to 99.5\% of total counts. Refer to caption of \autoref{MassAlpha005} for normalization and error panel. }
\label{MassAlpha010}
\end{figure}

\begin{figure}
\includegraphics[width=\columnwidth]{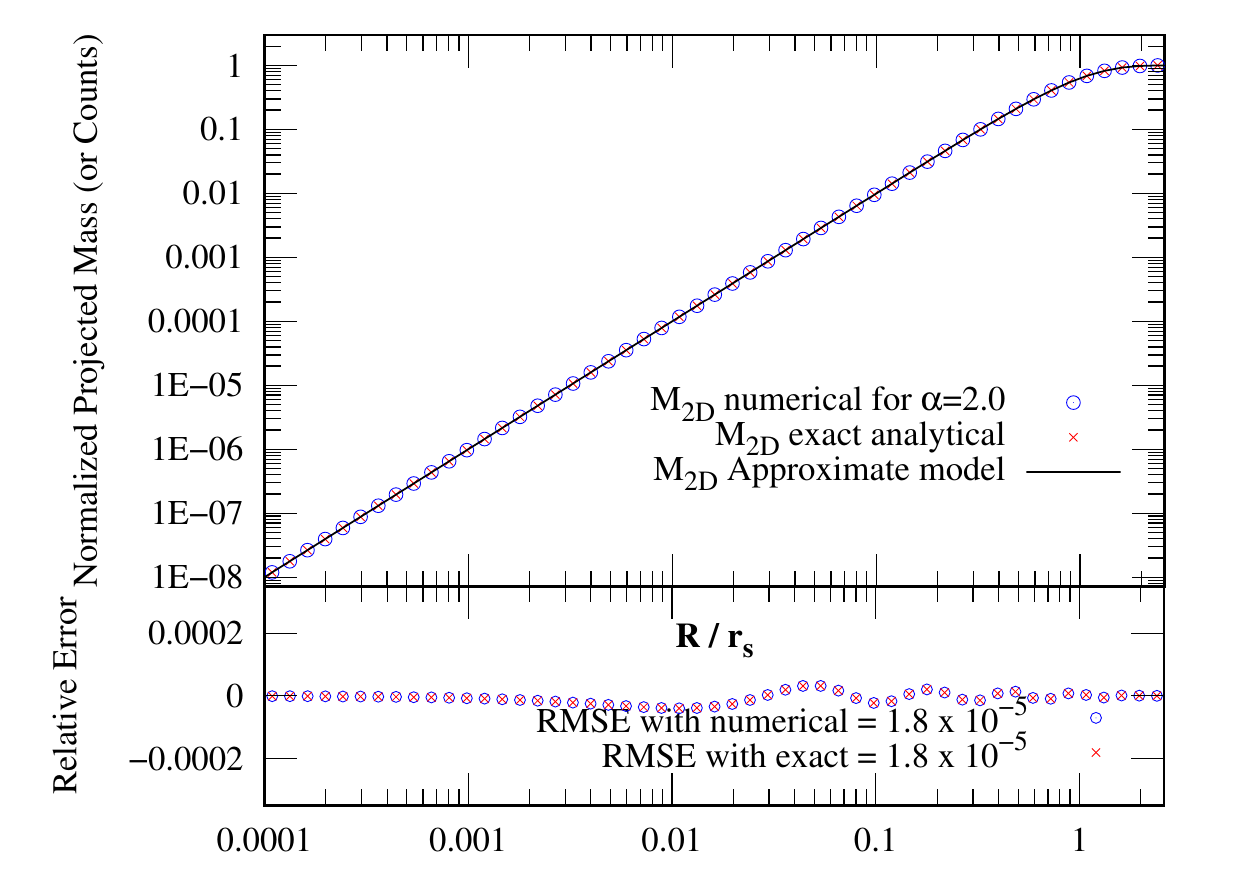}
\caption{Analytical model of normalized $M_{2D}(R)$ for an Einasto profile with $\alpha=2.00$ (The Gaussian function)---the only case for which an exact analytical solution exists. The spatial extent covers 99.5\% of total counts. Refer to caption of \autoref{MassAlpha005} for details on normalization and error panel.}
\label{MassAlpha2}
\end{figure}

\begin{figure}
\includegraphics[width=\columnwidth]{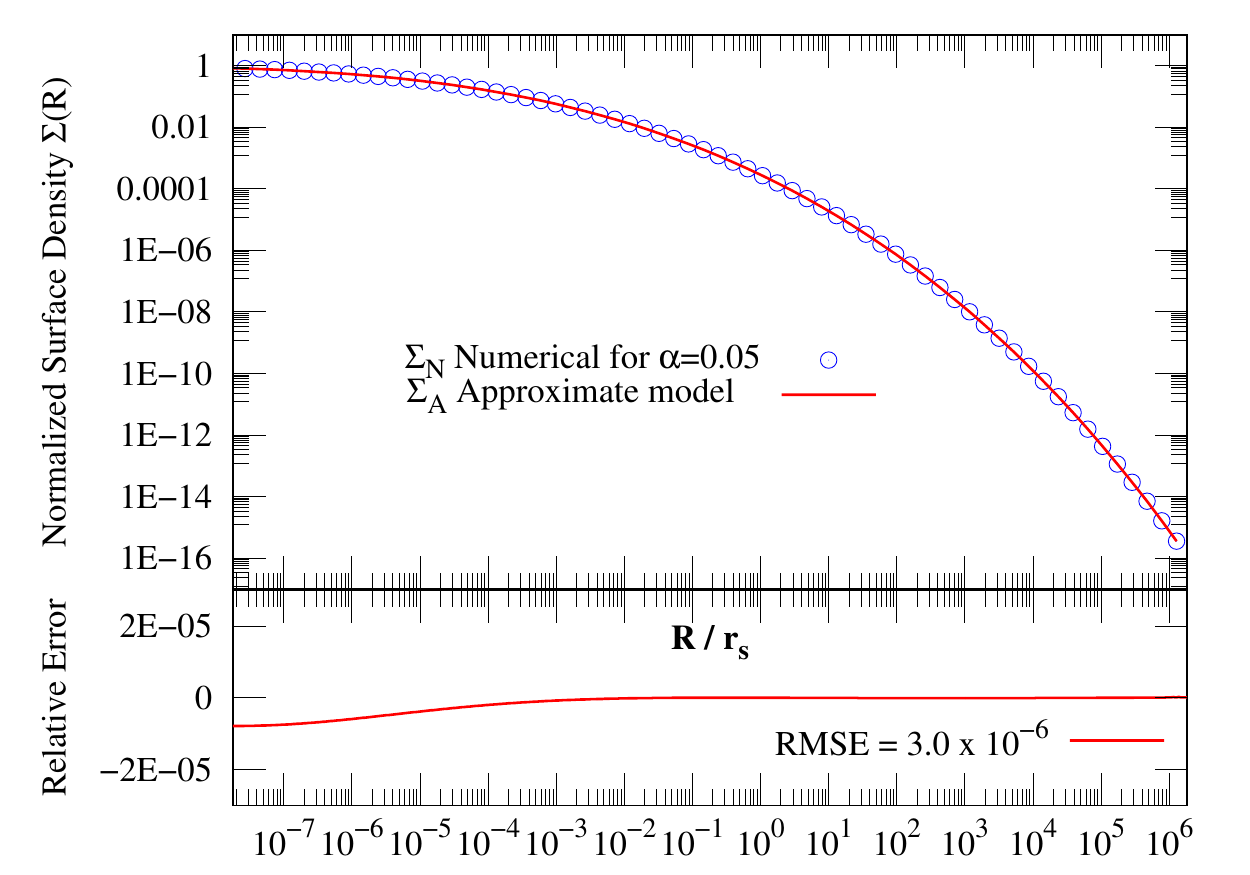}
\caption{Normalized $\Sigma (R)$ for an Einasto profile with $\alpha=0.05$. The high accuracy of the model spans 16 orders of magnitude in $\Sigma$ and 14 orders or magnitude in $R/r_s$---a spatial extent covering 99.5\% of total counts. Refer to caption of \autoref{MassAlpha005} for details on normalization and error panel.}
\label{SigAlpha05}
\end{figure}

\begin{figure}
\includegraphics[width=\columnwidth]{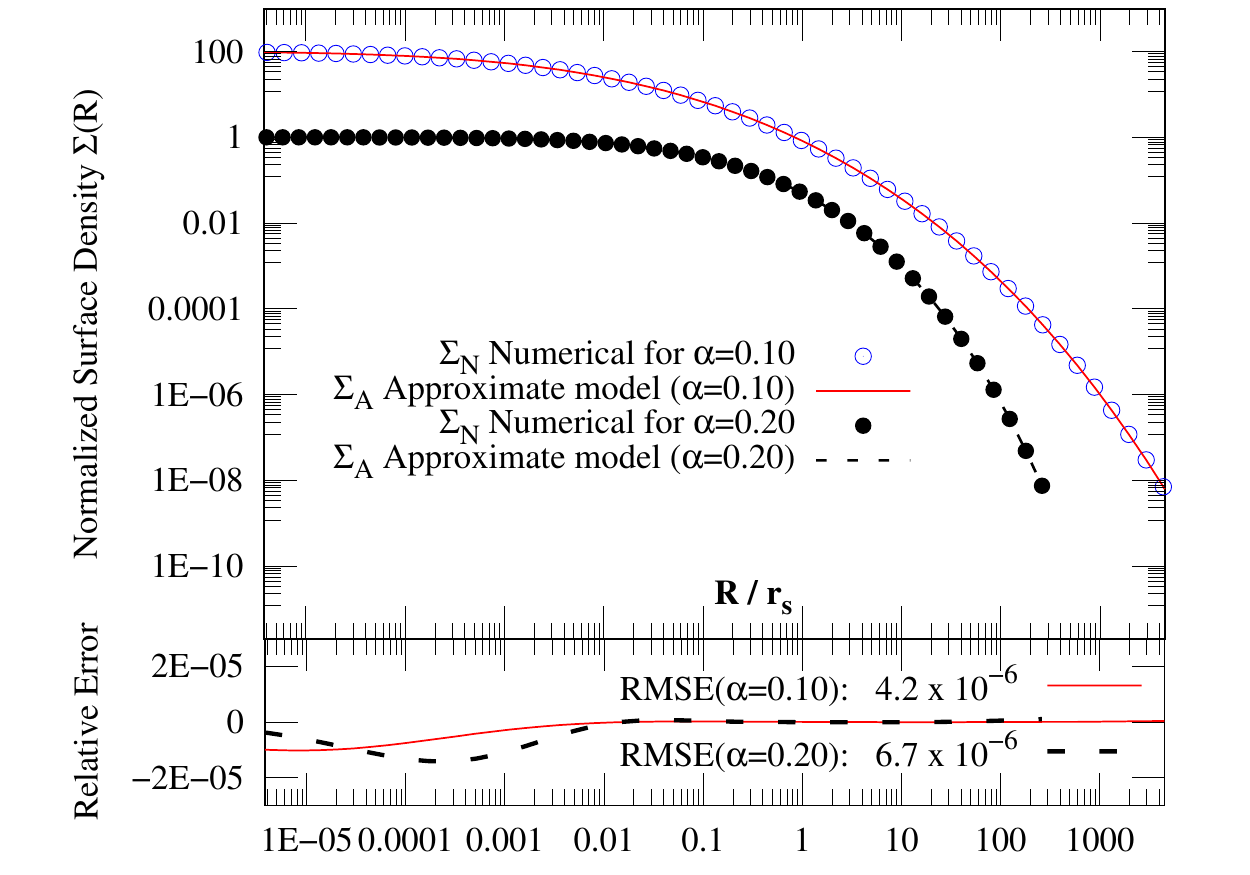}
\caption{Analytical models of normalized $\Sigma (R)$ for Einasto profiles with $\alpha=0.10$ (solid) and with $\alpha=0.2$ (dashed), the former scaled up by a factor of 100. The spatial extent covers 99.5\% of total counts. Refer to caption of \autoref{MassAlpha005} for details on normalization and error panel.}
\label{SigAlpha010}
\end{figure}

\begin{figure}
\includegraphics[width=\columnwidth]{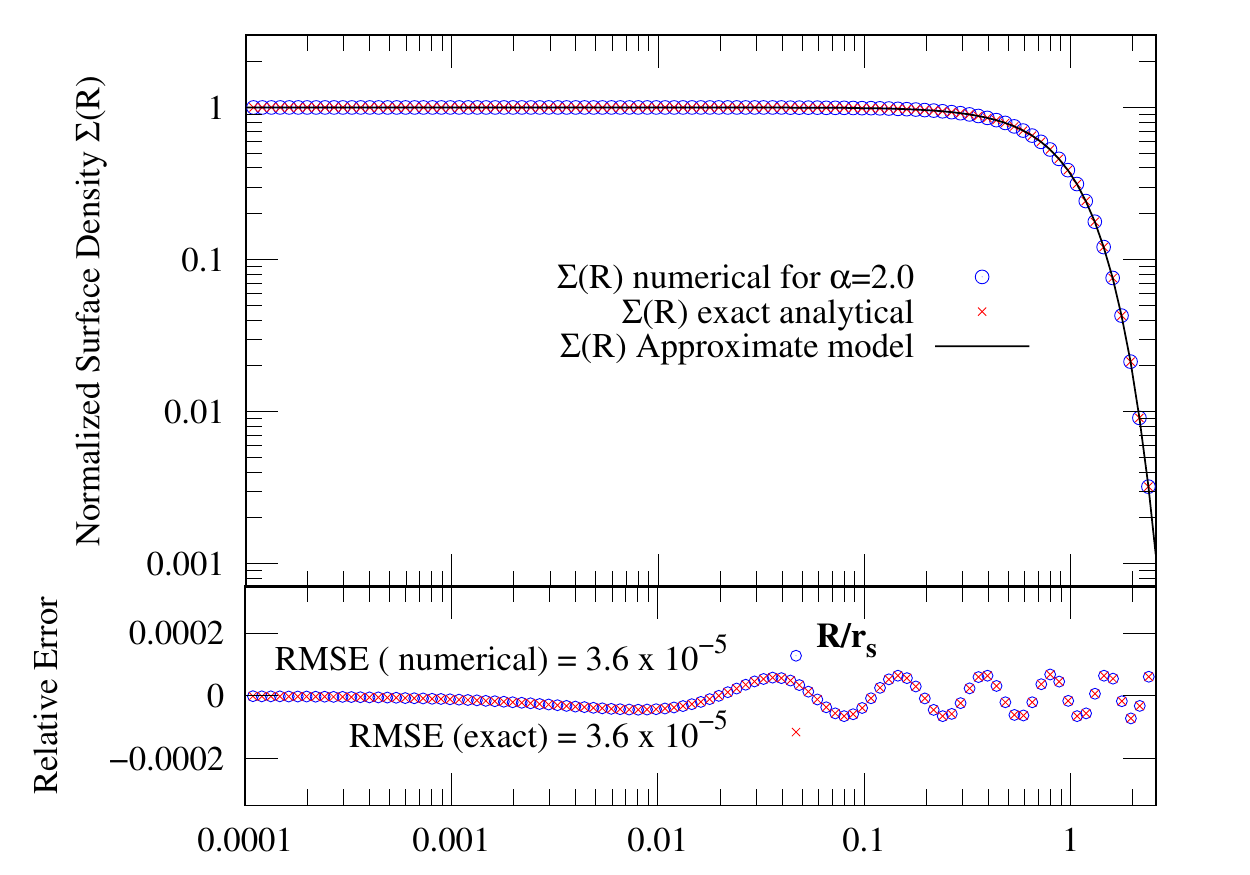}
\caption{Analytical model of normalized $\Sigma (R)$ for an Einasto profile with $\alpha=2.00$ (The Gaussian function)---the only case for which an exact analytical solution exists. The spatial extent covers 99.5\% of total counts. Refer to caption of \autoref{MassAlpha005} for details on normalization and error panel.}
\label{SigAlpha2}
\end{figure}

Considering that the $c_i$ and the $\zeta_i$ are independent of $\alpha$ (as per the discussion above), we observe:
\begin{enumerate}
\item that we can derive a self-consistent expression for the surface density $\Sigma(R)$ using $\Sigma(R)=\frac{1}{2\pi R} \frac{d M_{2D}(R)}{dR}$ to get:
\begin{multline}\label{Abelmodexp}
\Sigma (R)=\frac{2r_s \rho_0}{\alpha \ b^{1/\alpha}} \Bigg[\frac{\alpha \ b^{1/\alpha}}{2} \Big(\frac{R}{r_s}\Big) \sum_{i=0}^6 (c_i \zeta_i) \ \exp \Big(-b \Big(\frac{\zeta_i R}{r_s}\Big)^{\alpha}\Big) \\ 
+ 2 \ \Gamma \Big(\frac{1}{\alpha}, b \Big(\frac{R}{r_s}\Big)^{\alpha} \Big) - \sum_{i=0}^6 c_i \Gamma \Big(\frac{1}{\alpha}, b\Big(\frac{\zeta_i R}{r_s}\Big)^{\alpha}\Big) \Bigg]
\end{multline} 
---this is easier to see by first taking the derivative of \autoref{initial2}.
\item And, that we can estimate the $c_i$ and the $\zeta_i$ (assumed independent of $\alpha$) by fitting our model (\autoref{Abelmodexp}) to $\Sigma(R)$ for $\alpha=2$---the only $\alpha$ for which an exact analytical solution exists (\autoref{sig2exact}). It turns out that the parameters so estimated (see \autoref{coeff}), using a non-linear least squares Levenberg--Marquardt method, yield errors $\sim 10^{-5}$ for both $\Sigma(R)$ and $M_{2D}(R)$ (\autoref{fracerror}, Figs.\ref{MassAlpha005}-\ref{SigAlpha2}) for profiles of practical interest ($\alpha \lesssim 2.0$); and errors of $\sim 10^{-6}$ for $\alpha \lesssim 0.4$ which have been found to describe the density profiles of dark matter haloes in N-body simulations as well as the outer (most-dominant) component of massive ellipticals (\autoref{intro}).
\end{enumerate}
\begin{table}	
\begin{center}
\caption {Coefficients $c_i$ and $\zeta_i$ for $M_{2D}(R)$ \& $\Sigma(R)$}\label{coeff}	
\begin{tabular}{|l|l|l|}
\hline 
$i$ & $c_i$ & $\zeta_i$ \\
\hline
0 & \textcolor{magenta}{\bf 0.2955884256 \bf} & 1.0 \\
1 & 0.374929178 & 1.06186952774 \\
2 & 0.1912226623 & 1.2949052775 \\
3 &  0.0914083 & 1.848732123 \\
4 & 0.0352298967 & 3.180027854 \\
5 & 0.0099021431 & 6.78774063 \\
6 & 0.0017193943 & 21.2555914 \\
\hline
\end{tabular}
\begin{tablenotes}
Note: The correct value of $c_0$ is listed in this table. The published paper had an incorrect value corrected in an \href{http://dx.doi.org/10.1093/mnras/stab1760}{erratum} to the paper.
\end{tablenotes}
\end{center}
\end{table}
%
\begin{table}
\vspace{-3mm}
\begin{threeparttable}	
\caption {RMS of Relative Error for some values of $\alpha$}\label{fracerror}	
\begin{tabular}{|c|c|c|l|}
\hline 
$\alpha$ & Relative Error & Relative Error & Radial Extent (in $R/r_s$)\\ 
&for $M_{2D} (R)$ & for $\Sigma(R)$  & for 99.5\% of total mass\\
\hline
$0.05$ & $2.9 \times 10^{-6}$ & $3.0 \times 10^{-6}$ & $1,830,000$ \\
\hline
0.10 & $4.4 \times 10^{-6}$ &  $4.2 \times 10^{-6}$ & $4,150 $ \\
0.20 & $6.7 \times 10^{-6}$ & $6.7 \times 10^{-6}$  & $140$ \\
0.50 & $1.0 \times 10^{-5}$ & $1.0 \times 10^{-5}$ & $12.5$ \\
1.00 & $1.3 \times 10^{-5}$ & $1.4 \times 10^{-5}$ & $4.64$~~~~(the exponential) \\
2.00 & $1.8 \times 10^{-5}$ & $3.6 \times 10^{-5}$ & $2.53$~~~~(the gaussian) \\
\hline
3.00 & $3.0 \times 10^{-5}$ & $1.1 \times 10^{-4}$ & $2.00$ \\
5.00 & $7.4 \times 10^{-5}$ & $4.9 \times 10^{-4}$ & $1.60$ \\
\hline
\end{tabular}
\begin{tablenotes}
\item NOTE:-- Column 2 and Column 3 list the RMS of relative errors over a domain (column 4) encompassing $\sim 10^{-10}$\% to $99.5 \%$ (12 orders of magnitude) of the intrinsic total mass (or counts) $M_{3D total}$ (equation \ref{cum3Dtot}). The horizontal lines demarcate the domain of practical interest $0.1 \lesssim \alpha \lesssim 2.0$ described in \autoref{intro}. While the RMS errors for only some values of $\alpha$ are shown, this level of accuracy is maintained for all $\alpha \lesssim 5$. 
\end{tablenotes}
\end{threeparttable}
\end{table}

As a cross-check, we observe that \autoref{Ap2dcount} for $M_{2D}(R)$ $\to$  \autoref{cum3Dtot} for the total mass (counts) as $R \to \infty$ and \autoref{Abelmodexp} for $\Sigma(R)$ $\to$ \autoref{sigma0} for the projected central density as $R \to 0$. Further, the models under projection for $M_{2D}(R)$ and for $\Sigma(R)$ are in terms of the same three parameters describing the intrinsic volume density ($\rho_0, r_s, \alpha$) thereby giving us an insight into the 3D spatial density from the 2D models. Additionally, both models are specified in terms of exponential and gamma functions which are readily available in a wide range of computing platforms.
\vspace{-3mm}
\section{Summary \& Discussion}
\begin{figure}
\includegraphics[width=\columnwidth]{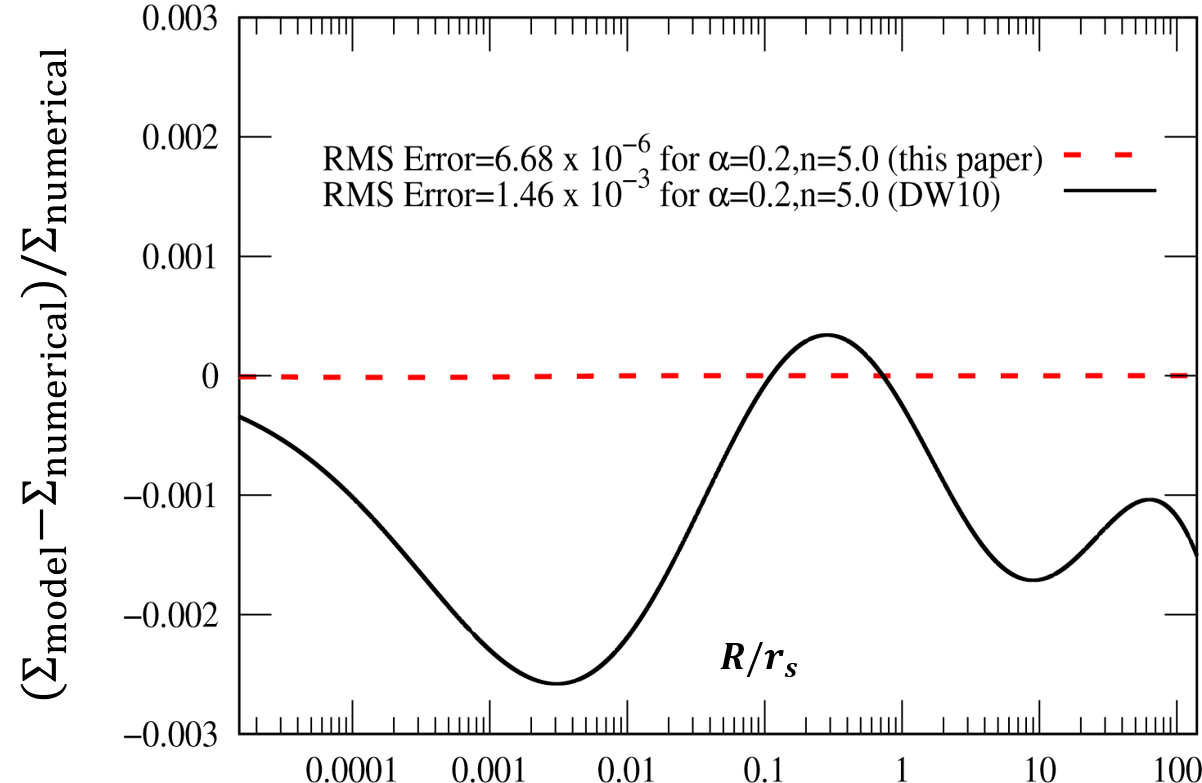}
\caption{A comparison of the relative error, with respect to the numerically integrated values, between the models for $\Sigma(R)$ presented in DW10 (solid) and in this paper (dashed) \autoref{Abelmodexp} for the case of $\alpha=0.2, n=5.0$.}
\label{comparemodels}
\end{figure}

In this paper, we have introduced a novel methodology (in \autoref{newmethod}) for developing an analytical approximation (\autoref{Ap2dcount}) for the projected mass (or counts), $M_{2D} (R)$, for a family of Einasto profiles with a shape parameter $0.05 \lesssim \alpha \lesssim 5.0$ (or Einasto index $0.2 \lesssim n \lesssim 20.0$)---a factor of 100. We have also seen (in \autoref{approxderiv}) how the expression for $M_{2D}(R)$ can be used to develop an analytical approximation for the projected (surface) density $\Sigma(R)$ (\autoref{Abelmodexp}). Both models are highly accurate for all $\alpha$ with fractional errors of $\sim 10^{-6}$ especially for $\alpha \lesssim 0.4$. This domain of $\alpha$ have been found to describe---i) the outer components of massive ellipticals that are believed to be dark matter dominated; and ii) the dark matter haloes in $\Lambda$CDM N-body simulations as described in \autoref{intro}. It is worth noting that while \citetalias{Dhar:surfden} have provided a fairly good ($\sim 10^{-3}$) approximation for $\Sigma(R)$, the result herein provides a significant reduction in the error in $\Sigma(R)$---by two to three orders of magnitude. To illustrate, we show one example in \autoref{comparemodels} for the case of $\alpha=0.2~(n=5.0)$, which is closest to the \citetalias{Nav04:einasto} model, where we can see that although the model in \citetalias{Dhar:surfden} is quite good, the model in this paper \autoref{Abelmodexp} is significantly better (with errors indistinguishable from zero); refer to \citetalias{Dhar:surfden} for errors for other $\alpha$.

The solutions presented in this paper for $M_{2D}(R)$ and for $\Sigma(R)$ (\autoref{Ap2dcount} and \autoref{Abelmodexp}) have broad applicability:
\begin{enumerate}
\vspace{-2mm}
\item For instance, when taken as the projected mass, \autoref{Ap2dcount} can be used to calculate several quantities due to gravitational lensing by circularly symmetric lenses characterized by Einasto profiles. This is because quantities such as the deflection angle $\alpha_E (R)$ and the mean convergence $\overline{\kappa}$ depend on $M_{2D}(R)$ through:
\begin{flalign}
\alpha_E (R)=\frac{4G \ M_{2D} (R)}{c^2 \ R}, \ and
\end{flalign}
\begin{flalign}\label{meankappa}
\overline{\kappa_E}(R)=\frac{M_{2D}(R)}{\Sigma_{cr} \pi R^2}
\end{flalign}
where, $\Sigma_{cr}$ is the critical surface mass density for lensing.

This allows us to determine the location of the tangential critical curve (from $\overline{\kappa_E}=1$) and the Einstein radius ($R_E$) for Einasto profiles
\begin{flalign}\label{Einrad}
by \ solving, \ ~~~~ M_{2D}(R_E)=\Sigma_{cr} \pi R_E^2
\end{flalign}
We can also use the analytical approximation for $\Sigma_E (R)$ (\autoref{Abelmodexp}) to calculate the convergence for Einasto profiles from
\begin{flalign}
 \kappa_E=\Sigma_E (R)/\Sigma_{cr} 
\end{flalign}
and along with the expression for $\overline{\kappa_E}$ (equation~\eqref{meankappa}), we can now calculate other quantities of interest in gravitational lensing due to a lens with an Einasto profile---such as, the shear ($\gamma_E$), the magnification ($\mu_E$) and the distortion ($q_E$) of lensed images of a circular source---using the relationships in \citet{Miralda:1991} :
 \begin{flalign}
\gamma_E &=\overline{\kappa_E}-\kappa_E \ , \\
\mu_E &=\Big[(1-\kappa_E)^2-(\overline{\kappa_E}-\kappa_E)^2\Big]^{-1} \ , \ \ and \\
q_E &=(1-\overline{\kappa_E})\Big[1-2\kappa_E+\overline{\kappa_E}\Big]^{-1} \, 
\end{flalign}
respectively. And the location of the radial critical curve from 
\begin{align}
1-2\kappa_E+\overline{\kappa_E}=0
\end{align}
The above discussion highlights the significance of \autoref{Ap2dcount} and \autoref{Abelmodexp} for $M_{2D}(R)$ and $\Sigma(R)$ -- that they allow us to analytically calculate several quantities in gravitational lensing (due to lenses characterized by an Einasto profile), and in terms of functions in standard math libraries which has not been possible until now.

\item Likewise, when $\rho(r)$ is taken as a luminosity density or a number density, \autoref{Ap2dcount} allows us to easily calculate the projected luminosity and the columnar number counts,  while \autoref{Abelmodexp} allows us to calculate the surface brightness. It is worth noting that both expressions are in terms of the same three parameters ($\rho_0, r_s$ and $\alpha$) defining the intrinsic volume density thereby giving us an insight into the 3D density distribution.

It is instructive to note that this new methodology of first developing an expression for the projected mass (as in \autoref{newmethod}) and of then calculating the projected surface density $\Sigma(R)$ (as in \autoref{approxderiv}), can be applied to other profiles so long as the integrals of both $\rho(r)$ and $r^2 \rho(r)$ are bounded and can be calculated analytically ---that is, for profiles with finite total mass (or counts).
\end{enumerate}
%
\vspace{-3mm}
\section*{Acknowledgements}
I thank Liliya Williams and the anonymous reviewer for their suggestions that helped improve the presentation of this work.
\vspace{-3mm}
\section*{Data Availability}
No new data were generated or analysed in support of this research.
\vspace{-2mm}
\bibliographystyle{mnras}
\bibliography{2020papers_short2} 
\bsp	
\label{lastpage}
\end{document}